\documentclass[runningheads]{llncs}

\usepackage[mobile]{eccv}

\usepackage{eccvabbrv}

\usepackage{graphicx}
\usepackage{booktabs}
\usepackage{multirow}
\usepackage{algorithm}
\usepackage{algpseudocode}

\usepackage[accsupp]{axessibility}  
\usepackage{caption}

\usepackage[hypertexnames=false]{hyperref}

\usepackage{orcidlink}

\usepackage{multibib}
\newcites{supp}{Supplementary References}
\usepackage{xr}
\begin{document}

\title{NaP-Control: Navigating Diffusion Prior for Versatile and Fast Character Control} 

\titlerunning{NaP-Control}

\author{}
\author{Chia-Wen Chen\orcidlink{0009-0005-9290-1379} \and
Yan Wu\orcidlink{0009-0007-1260-0045} \and
Korrawe Karunratanakul\orcidlink{0000-0002-0204-132X} \and
Siyu Tang\orcidlink{0000-0002-1015-4770}}

\authorrunning{C.-W.~Chen et al.}

\institute{}
\institute{ETH Zurich\\
\email{chiachen@ethz.ch, \{yan.wu, korrawe.karunratanakul, siyu.tang\}@inf.ethz.ch} \\
\url{https://chiawenchen.github.io/nap-control-project/}}
\maketitle

\begin{center}
  \newcommand{\teaserwidth}{\textwidth}
  \centerline{\includegraphics[width=\linewidth]{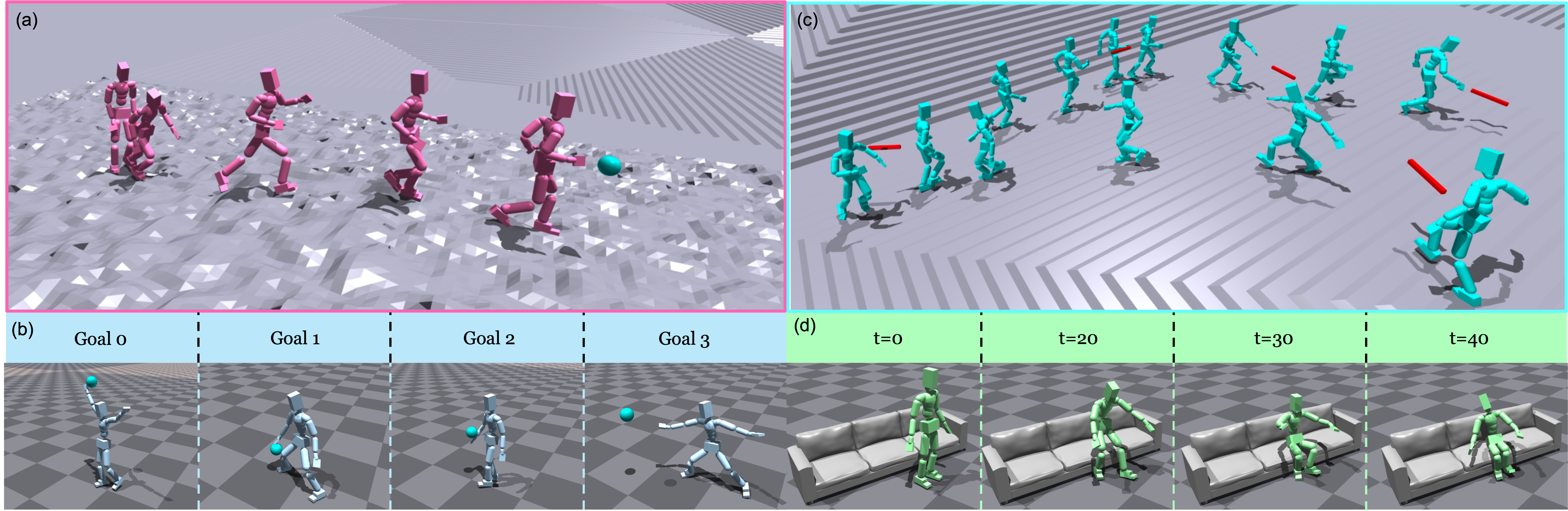}}
    \captionof{figure}{NaP-Control is a latent noise optimization framework combining reinforcement learning and diffusion-based prior for physics-based character control. We showcase its effectiveness in (a) far goal reaching, (b) agile hand reaching, (c) velocity control, (d) object interaction tasks, as well as its adaptation on uneven terrains.
    }
    \label{fig:teaser}
\end{center}

\begin{abstract}
Achieving precise, versatile whole-body character control in physics-based animation remains challenging. Recent diffusion-based policies generate rich and expressive motions but typically rely on gradient-based test-time guidance to satisfy task objectives, which is slow and can reduce robustness. We introduce \textbf{NaP}-\textbf{C}ontrol (\textbf{Na}vigating Diffusion \textbf{P}rior for Versatile and Fast Character \textbf{C}ontrol), abbreviated as \textbf{NaP}. Our method uses reinforcement learning to manipulate the latent noise of a task-agnostic diffusion policy prior, steering it toward task-specific behaviors for fast, robust control with high motion fidelity. In contrast to methods that rely solely on offline training, NaP interacts with the environment during training to correct motions and optimize task rewards, improving success rates and enabling adaptation to challenging scenarios. By directly predicting task-optimized diffusion noise, NaP eliminates iterative guidance during denoising and enables efficient inference. Experiments show that NaP attains higher success rates and faster inference while preserving natural motion across diverse tasks.
  \keywords{Physics-based Character Control \and Diffusion Models \and Reinforcement Learning}
\end{abstract}
\section{Introduction}
\label{sec:intro}
Physics-based whole-body character control~\cite{deepmimic, phc, pulse, maskedmimic, uniphys, huang2025diffuse, maskedmanipulator, xiao2023unified, yao2022controlvae, yao2024moconvq} aims to generate physically plausible human motion that remains robust and responsive under a wide range of control inputs and in complex environments. A central challenge is to produce high-quality behaviors that both satisfy physical constraints and adapt efficiently to different task objectives.

A prominent line of work tackles this challenge by learning reusable motion priors from large-scale motion datasets. In particular, VAE-based action priors~\cite{pulse, yao2022controlvae, maskedmimic} allow reinforcement learning (RL) to operate in a compact latent space, but their stateless, single-step formulation can limit temporal consistency and overall expressiveness.
More recently, diffusion-based policies~\cite{uniphys,huang2025diffuse,beyondmimic} have advanced physics-based character control by modeling long-horizon motion distributions, yielding smoother and more expressive behaviors. However, these approaches are generally trained purely offline without interacting with the environment during learning, which constrains their robustness for downstream control tasks. To incorporate task objectives, prior work \cite{beyondmimic, uniphys} typically employs loss-based guidance at test time, which requires differentiable objectives and iterative optimization during inference. This results in limited applicability and substantial computational overhead. Consequently, how to effectively leverage expressive diffusion policy priors for robust and efficient task adaptation in physics-based character control remains largely an open question.

To address this challenge, we introduce NaP-Control, a framework that adapts a pretrained task-agnostic diffusion policy by learning to navigate its input noise space through RL. Prior work has investigated diffusion-noise optimization for behavioral control, including RL-based steering in robotic manipulation~\cite{steering} and direct noise optimization for kinematics-based motion synthesis~\cite{dno}. Our method extends this emerging paradigm to a substantially more challenging setting: physics-based whole-body character control, in which task performance depends on closed-loop interaction with the environment, contact-rich dynamics, and long-horizon stability. In this setting, small perturbations can induce loss of balance, rendering purely offline methods inadequate. We therefore treat the pretrained diffusion policy as an expressive action manifold and train an RL policy to generate task-aware latent noise through direct interaction with the environment. This formulation enables the controller to adapt to environmental feedback, optimize task-specific rewards, and accommodate tasks that lack simple differentiable objective functions, while preserving the naturalness and smoothness encoded by the diffusion prior. Notably, NaP-Control eliminates the need for both extensive reward engineering, which is typically required in reinforcement learning, and iterative gradient-based guidance at inference time, thereby achieving substantially higher computational efficiency than existing diffusion-based control frameworks.

Our contributions are summarized as follows:

(1) We introduce NaP-Control, a reinforcement learning framework that adapts pretrained diffusion-based action priors for task-driven physics-based character control by producing the initial diffusion noise.

(2) We show that navigating the initial diffusion noise space provides an effective way to incorporate environmental feedback into diffusion-based control, overcoming the limitations of purely offline training and expensive test-time guidance while preserving motion fidelity.

(3) We demonstrate that NaP-Control achieves faster inference and higher success rates across diverse locomotion tasks, including challenging traversal on unseen terrains, compared to diffusion-based policies while maintaining natural and human-like whole-body movement in all tasks.

\section{Related Work}
\subsection{Kinematics-based Human Motion Generation}
Human motion generation aims to synthesize realistic motion sequences, typically represented as trajectories of joint rotations and global root motion over time. A large body of prior work focuses on kinematics-based generation, which emphasizes visual realism and diversity without explicitly modeling physical dynamics. These methods have achieved strong results in multi-modality conditioned human motion generation~\cite{mdm, motiondiffuse, unimogen, anytop, wang2025fg, li2025genmo, yang2025unimumo}, and controllable motion editing and synthesis~\cite{gmd, dno, pinyoanuntapong2025maskcontrol, li2025simmotionedit}.

A central direction in this area is controllable motion synthesis. Existing approaches~\cite{dno, pinyoanuntapong2025maskcontrol,TLcontrol,huang2024controllable} introduce control through target trajectories, obstacle constraints, or joint-level spatio-temporal specifications. In diffusion-based methods, such control is often incorporated through guidance~\cite{gmd,omnicontrol,shi2024interactive,dai2024motionlcm} or noise optimization~\cite{dno}, while other methods allow more fine-grained editing of specific body parts~\cite{mmm} or motion segments~\cite{pinyoanuntapong2025maskcontrol}. More recent work has also extended kinematics-based generation to online settings~\cite{dart, motion_streamer, closd}, where motion is produced autoregressively or in a streaming manner based on motion history and language input, enabling more responsive and continuous generation.

Despite their strong visual quality, kinematics-based methods do not explicitly enforce physical consistency, and thus often suffer from artifacts such as floating, foot skating, ground penetration, and jitter~\cite{physdiff}. These limitations make them difficult to deploy in embodied settings and motivate physics-based motion generation, which incorporates dynamics and contact interactions to produce physically feasible behaviors.

\subsection{Physics-based Character Control}

Early work in physics-based character control trains task-specific controllers using reinforcement learning~\cite{deepmimic, de2010feature, liu2010sampling}. To improve motion realism, later methods incorporate motion capture data through imitation and adversarial learning~\cite{amp, ase, calm}. While effective, these approaches typically require separate policies for different tasks, limiting scalability and generalization.
Recent work learns generalized motion tracking controllers or reusable action priors from large-scale motion datasets~\cite{phc, maskedmimic, maskedmanipulator, pacer+, beyondmimic, yao2022controlvae, yao2024moconvq}.
For example, PHC~\cite{phc} and MaskedMimic~\cite{maskedmimic} show that a single controller can track diverse motions and support multiple tasks. PULSE~\cite{pulse} distills a VAE-based action prior from motion data to enable efficient downstream learning, while MaskedMimic~\cite{maskedmimic} learns a multi-task controller. Nevertheless, VAE-based priors and single-step formulations often limit motion expressiveness and temporal coherence.
More recently, diffusion models have significantly improved motion modeling and have been applied to physics-based control~\cite{uniphys, truong2024pdp, closd, huang2025diffuse}. CLoSD~\cite{closd} integrates a diffusion-driven kinematic generator with a physics-based tracker. Other methods, including UniPhys~\cite{uniphys} and DiffuseCLoC~\cite{huang2025diffuse}, distill diffusion policies from offline datasets and adapt them to new tasks using test-time guidance. Yet, these techniques rely on iterative, gradient-based optimization during inference, resulting in slow and sometimes unstable control and limiting adaptation due to the lack of direct environment interaction.

In contrast, we build upon a pretrained diffusion policy and propose an efficient framework to leverage it as a temporally smooth, expressive action prior. By learning to navigate the latent noise space using task rewards and online environment feedback, our method enables robust, goal-directed behavior without costly test-time guidance, while supporting a wide range of downstream tasks with substantially faster inference.

\subsection{Improving Diffusion Model with Reinforcement Learning}
Reinforcement learning (RL) has recently been explored as an effective way to improve and adapt pretrained models across domains such as image generation~\cite{fan2023dpok, fan2023optimizing, black2023training}, language modeling~\cite{bai2022training, liuchain}, and robotics~\cite{dppo, steering}. The goal is to align generative outputs with downstream objectives, such as human preference and physical feasibility, which are difficult to capture through supervised learning alone. One approach directly fine-tunes diffusion models using reward signals~\cite{christiano2017deep, black2023training, zhang2024large, dppo}.
Examples include RLHF-style methods~\cite{christiano2017deep} for improving image generation and DPPO~\cite{dppo} and IDQL~\cite{hansen2023idql} for fine-tuning diffusion policies in robotics. However, while effective, joint optimization is often unstable and sample-inefficient. Another direction leverages RL without modifying the pretrained diffusion model by learning an auxiliary policy, such as learning a residual policy~\cite{sentinel} on top of the frozen prior, though these methods mainly provide local corrections. More recently, latent noise steering~\cite{eyring2024reno, steering} has been proposed, where RL controls the input noise of the diffusion process, and DSRL~\cite{steering} shows that noise steering can effectively control diffusion-based policies for low-dimensional robotic manipulation. In contrast to prior work that primarily targets low-dimensional or kinematic settings, we scale latent noise steering to high-dimensional, physics-based whole-body control.

\section{Whole-Body Control via Diffusion Noise Navigation}
\begin{figure}[tb]
  \centering
  \includegraphics[width=\linewidth]{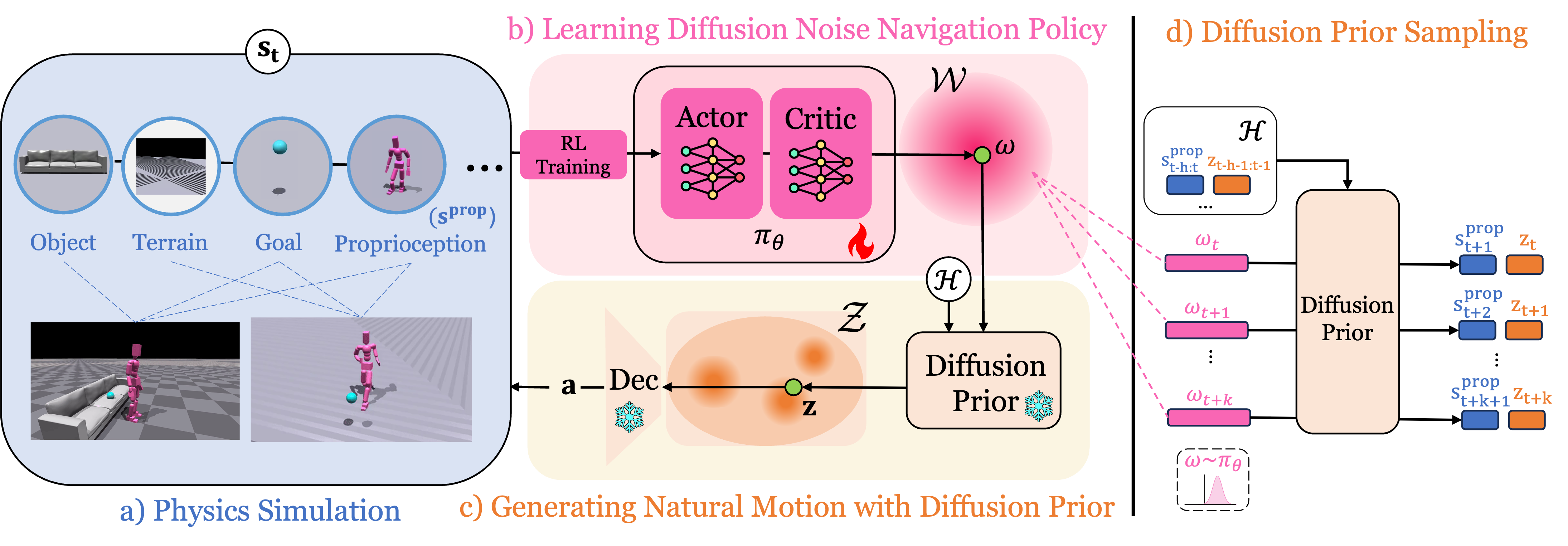}
  \caption{Framework overview. (a) The RL policy $\pi_{\theta}$ receives environment and proprioceptive states from the simulation. (b) The actor learns to predict optimal noise $\boldsymbol{\omega} \in \mathcal{W}$ aligned with task goals. (c) These predicted noises are denoised and decoded into executable actions $\mathbf{a}$ via a pretrained diffusion prior and a latent action decoder. Resulting transitions are then used to iteratively optimize the noise navigation policy with RL. (d) As a natural motion manifold, the diffusion prior can map predicted noises to future states $\mathbf{s}^{\text{prop}}$ and latent actions $\mathbf{z}$ conditioned on the trajectory history $\mathcal{H}$.}

  \label{fig:framework}
\end{figure}
To facilitate the generation of natural motions that satisfy task objectives, we propose a framework that utilizes a pre-trained diffusion model as a generative action prior and steers it toward task-specific control. We hypothesize that a task-agnostic diffusion policy provides an expressive movement prior.
By searching within this prior space, a wide range of downstream tasks can be achieved while preserving motion smoothness and naturalness, which are inherently regularized by the action prior. We therefore pretrain a diffusion-based action prior (Section~\ref{sec:method_diffusion_prior}) and demonstrate that optimizing the initial noise of the action prior through RL (Section~\ref{sec:method_noise_steer}) allows for the effective learning of task-specific policies (Section~\ref{sec:method_task}). Fig.~\ref{fig:framework} provides an overview of the overall framework.

\subsection{Diffusion-based Action Prior Pretraining}
\label{sec:method_diffusion_prior}
We first pretrain a diffusion-based behavior generative model that serves as a task-agnostic action prior, grounding policy rollouts within a high-fidelity motion manifold. This prior captures the distribution of natural humanoid behaviors and later acts as a strong regularizer for downstream task learning.
We leverage a large-scale state–action dataset constructed by tracking reference human motions from AMASS~\cite{amass} in a physics simulator. Following UniPhys~\cite{uniphys}, the diffusion prior is implemented as a causal transformer-based diffusion model that conditions on the history of previous states and actions. Given this context, the model iteratively denoises latent noise to generate a chunk of future actions and predicted states, as shown in Fig.~\ref{fig:framework}(d). We adopt their training paradigm while introducing a more compact state representation that removes redundant kinematic features.
Specifically, UniPhys \cite{uniphys} employs an expansive state representation that contains a variety of kinematic features, including joint positions, velocities, and other redundant quantities. In contrast, we retain only the essential dynamic features $\mathbf{S} = (\mathbf{r}_{1:T}, \mathbf{q}_{1:T}, \mathbf{w}_{1:T})$, which include:
\begin{itemize}
    \item Global Root Trajectory ($\mathbf{r}_{1:T}$): The root trajectory is canonicalized relative to the first-frame coordinate system, where the initial position is at the origin facing the positive $y$-axis. It includes the root position $\zeta_t \in \mathbb{R}^3$, 6D orientation $\phi_t \in \mathbb{R}^6$, linear velocity $\dot{\zeta}_t \in \mathbb{R}^3$, and angular velocity $\dot{\phi}_t \in \mathbb{R}^3$.
    
    \item Local Joint Features: Joint-level data is canonicalized to a per-frame local coordinate system centered at the pelvis projection on the ground. This includes 6D joint rotations $\mathbf{q}_t \in \mathbb{R}^{J \times 6}$, and angular velocities $\mathbf{w}_t \in \mathbb{R}^{J \times 3}$, and exclude local joint positions $\mathbf{p}_t \in \mathbb{R}^{J \times 3}$ and velocities $\mathbf{v}_t \in \mathbb{R}^{J \times 3}$.
\end{itemize}
This compact representation maintains motion fidelity and robustness while lowering the dimensionality of the learning space. Consequently, both diffusion-prior pretraining and the following reinforcement learning are more efficient.

We model the joint state-action distribution with the diffusion prior following previous works~\cite{uniphys, huang2025diffuse}. To manage the high dimensionality of the raw action space $\mathcal{A}$, we use a representation from PULSE~\cite{pulse} to first encode the raw action into a compact latent action space $\mathcal{Z}$. 
The diffusion prior therefore models the trajectory of state–latent pairs $\mathbf{x}_t = (\mathbf{s}_t, \mathbf{z}_t)$. During inference, the model conditions on the historical trajectory $\mathcal{H}$ and iteratively denoises a sequence of future latent actions, which are then decoded into executable control signals via the pretrained action decoder $\mathbf{a_t} = \mathcal{D}(\mathbf{s}_t, \mathbf{z}_t)$. 


\subsection{Navigating Diffusion Noise for Whole-Body Control}
\label{sec:method_noise_steer}

To navigate the manifold of the pre-trained diffusion prior and satisfy downstream task objectives, we learn a latent noise navigation policy via RL to select the initial diffuse noise. This navigation mechanism is critical, as the initial noise determines the specific trajectory the diffusion process takes through the learned motion space \cite{dno, initno}. By treating the noise space as a controllable search manifold, we can effectively guide the generative process toward physically plausible and task-specific behaviors.

\subsubsection{Navigation Mechanism in Natural Motion Prior.} 
As illustrated in Fig.~\ref{fig:framework}a, the latent noise navigation policy $\pi_\theta$ processes the observation $\mathbf{s}_t = \{\mathbf{s}^{\text{prop}}, \mathbf{s}^{\text{env}}\}$, where $\mathbf{s}^{\text{prop}}$ denotes the character's current proprioceptive state and $\mathbf{s}^{\text{env}}$ encodes task-specific environment state. Unlike standard controllers that map states directly to raw joint actions $\pi(\mathbf{a} \mid \mathbf{s})$, our navigation policy $\pi_\theta$, following the formulation of DSRL \cite{steering}, operates on the latent noise manifold of a pretrained diffusion prior. To preserve the diffusion prior’s sequential generation and temporal consistency while maintaining a low-dimensional search space for efficient RL exploration, we duplicate the latent noise $\boldsymbol{\omega}_t$ into a $k$-step noise chunk \cite{steering}. The predicted noise chunk $\boldsymbol{\omega}_{t:t+k-1}$ is further denoised into a task-aligned latent action chunk $\mathbf{z}_{t:t+k-1}$ using the DDIM-ODE solver \cite{Song2021-ac}:

\begin{align}
\label{eq:noise_steering}
{\boldsymbol\omega_t} &= \pi_\theta(\mathbf{s}_t) \\
\mathbf{z}_{t:t+k-1} &= \mathcal{M}(\text{repeat}(\boldsymbol{\omega}_t, k) | \mathcal{H})
\end{align}

The latent actions are then decoded to joint-level raw actions $\mathbf{a}_{t:t+k-1}$ via the frozen action decoder $\mathcal{D}$ \cite{pulse}. The joint-level actions are executed open-loop, i.e., without per-step feedback, within the physics simulator, which returns the subsequent proprioceptive $\mathbf{s}_{t+1:t+k}$ and environmental observations. We encapsulate both the diffusion prior as well as the latent action decoder as part of the Markov Decision Process (MDP) environment for our latent noise navigation policy, as shown in Fig.~\ref{fig:framework}(c). This architecture is crucial for grounding the final actions within a natural, human-like motion manifold. By doing so, it ensures physical plausibility and prevents the character from drifting into out-of-distribution states, a common failure mode in traditional test-time guidance methods.
\begin{algorithm}[H]
\caption{Latent Noise Navigation Policy Training}\label{alg:steering}
\begin{algorithmic}[l]
\Statex \textbf{Input:} Frozen diffusion prior $\mathcal{M}$, frozen decoder $\mathcal{D}$, physics simulator $\mathcal{SIM}$,
\Statex \hskip1.5em reward function $r(\cdot)$, terrain heightmap $H$ (optional), action chunk size $k$, 
\Statex \hskip1.5em initial proprioceptive state $\mathbf{s}^{prop}$ and environment states $\mathbf{s}^{env}$, horizon length $T$
\Statex \textbf{Output:} Noise optimizing policy $\pi_\theta$
\State Initialize policy $\pi_\theta$, experience buffer $\mathcal{B}$, and diffusion history buffer $\mathcal{H}$

\For{epoch = 1, 2, \dots, N}
    \State Reset environment; obtain $\mathbf{s}^{prop}$ and $\mathbf{s}^{env}$ from $\mathcal{SIM}$
    \State Reset diffusion history buffer $\mathcal{H}$
    
    \For{timestep $t = 0, k, 2k, \dots, T-k$}
        \State $\mathbf{s}_t \gets \{\mathbf{s}^{prop}_t, \mathbf{s}^{env}_t, \mathcal{H} \} $ 
        \State Sample latent noise $\boldsymbol{\omega} \sim \pi_{\theta_{old}}(\boldsymbol{\omega} \mid \mathbf{s}_t)$ 
        
        \State Denoise to latent action chunk: $\mathbf{z}_{t:t+k-1} = \mathcal{M}(\text{repeat}(\boldsymbol{\omega}, k) \mid \mathcal{H})$
        
        \State $R_{chunk} \gets 0$
        \For{$i = 0$ \textbf{to} $k-1$}
            \State $\mathbf{a}_{t+i} = \mathcal{D}(\mathbf{s}_{t+i}, \mathbf{z}_{t+i})$
            \State Execute $\mathbf{a}_{t+i}$ in $\mathcal{SIM}$, observe $\mathbf{s}_{t+i+1}$
            \State Compute reward $r_{t+i+1} = r(\mathbf{s}_{t+i+1})$
            \State Update $\mathcal{H} \gets \{\mathbf{s}_{t+i+1}, \mathbf{z}_{t+i}\}$
            \State $R_{chunk} \gets R_{chunk} + \gamma^i r_{t+i+1}$
        \EndFor
        \State Store transition $(\mathbf{s}_t, \boldsymbol{\omega}, R_{chunk}, \mathbf{s}_{t+k})$ in $\mathcal{B}$
    \EndFor
    
\State Compute advantage estimates $\hat{A}_t$ based on $\mathcal{B}$, for $t \in \{0, k, 2k, \dots, T-k\}$
    \State Optimize surrogate loss L wrt $\theta$ (m miniepochs, minibatch size b)
    \State $\pi_{\theta_{old}} \gets \pi_\theta$
\EndFor
\State \textbf{return} $\pi_\theta$
\end{algorithmic}
\end{algorithm}
\subsubsection{Latent Noise Navigation Policy Training.} 
To learn the latent noise navigation policy, our framework employs on-policy Reinforcement Learning (RL) via Proximal Policy Optimization (PPO) \cite{ppo}, shown in Fig.~\ref{fig:framework}b. During training, the policy iteratively shifts from producing stochastic noise with low task-efficacy toward generating noise distributions optimized for task fulfillment. We optimize actor and critic networks using the PPO clipped objective \cite{ppo} and Generalized Advantage Estimation (GAE) \cite{schulman2015high} to ensure training stability. We also incorporate a bounding loss on the policy's output, ensuring the generated noise remains within a stable value range. We treat the entire action chunk as a single micro-step in an MDP. Accordingly, rewards $r_i$ are calculated at each simulation step, but accumulated through a discount factor $\gamma$ to form the total chunk reward. The resulting transition tuple $(\mathbf{s}_t, \boldsymbol{\omega}, \sum_{i=1}^{k} \gamma^{i-1} r_{t+i}, \mathbf{s}_{t+k})$ is stored in the experience buffer for policy updates and $\mathbf{s}_{t+1:t+k}$ is saved to diffusion prior's history buffer $\mathcal{H}$ to provide temporal consistency for the next roll out. The complete training algorithm is shown in Algorithm \ref{alg:steering}.

In this manner, our framework uses the diffusion prior as a task-agnostic natural motion manifold, exploiting its expressive motion space to generate high-fidelity actions that both preserve motion realism and optimize downstream task rewards. By directly interacting with the environment during training, the framework adapts the pretrained prior to handle scenarios beyond the original motion distribution, including uneven terrain and object interaction. By replacing the standard stochastic Gaussian sampling $\boldsymbol{\omega} \sim \mathcal{N}(\mathbf{0}, \mathbf{I})$ used in traditional diffusion guidance with the deterministic noise output of our trained policy $\pi_\theta$, we eliminates the reliance on computationally expensive test-time guidance, enabling more efficient goal-conditioned task execution. In summary, the latent noise navigation policy together with the diffusion prior effectively bridges high-level task constraints with the low-level natural motion manifold, enabling robust, fast and versatile whole-body control.

\subsection{Applications}
\label{sec:method_task}
We demonstrate that our proposed framework is effective across various locomotion tasks while maintaining the intrinsic naturalness of the underlying motion manifold. Moreover, it exhibits robust performance in extended and complex scenarios, including object interaction and uneven terrain traversal, which are typically challenging for previous guidance-based diffusion policies~\cite{uniphys, huang2025diffuse}.


\subsubsection{Far Goal Reaching.}
The agent is tasked with reaching a target position $p_{\text{goal}}$, as illustrated in Figure~\ref{fig:teaser}(a). The reward consists of three terms: (1) a location reward that drives the agent toward the goal; (2) an orientation reward that encourages the agent to face the goal during approach; and (3) a stability penalty that promotes controlled, steady arrival. The stability term is crucial for suppressing unnatural orbiting behavior, where the agent circles around the goal to preserve momentum instead of settling at $p_{\text{goal}}$. More concretely,

\begin{equation}
\label{eq:r_reach}
R_{\text{reach}} = (1 - \lambda_1 - \lambda_2 ) \cdot R_{\text{location}} + \lambda_1 \cdot R_{\text{orientation}} + \lambda_2 \cdot R_{\text{stability}}
\end{equation}

\begin{equation}
\label{eq:r_location}
R_{\text{location}} = \exp(-\alpha_1 \cdot \|p_\text{{goal}} - p_\text{{joint}}\|_2)
\end{equation}

\begin{equation}
\label{eq:r_ori}
R_{\text{orientation}} = 
    \begin{cases} 
    1.0 & \text{if } \|p_\text{{goal}} - p_\text{{joint}}\|_2 \leq 0.3\,\text{m} \\
    \exp\left(\alpha_2 \cdot (\mathbf{d}_{\text{fwd}} \cdot \mathbf{d}_{\text{goal}} - 1.0)\right) & \text{otherwise}
    \end{cases}
\end{equation}

\begin{equation}
\label{eq:r_stab}
R_{\text{stability}} = 
    \begin{cases} 
    -\beta_1 \cdot (c_1 v + c_2 \omega) & \text{if } \|p_\text{{goal}} - p_\text{{joint}}\|_2 < 1.2\,\text{m} \\
    -\beta_2 \cdot (c_1 v + c_2 \omega) & \text{if } 1.2\,\text{m} \leq \|p_\text{{goal}} - p_\text{{joint}}\|_2 < 6.0\,\text{m} \\
    0 & \text{if } \|p_\text{{goal}} - p_\text{{joint}}\|_2 \geq 6.0\,\text{m}
    \end{cases}
\end{equation}

\subsubsection{Agile Hand Reaching.}
We further show that our framework is well-suited for tasks that demand high agility and precise end-effector control. We design an agile hand-reaching task (Fig.~\ref{fig:teaser}(b)) in which the agent must move its right hand to targets distributed throughout its reachable workspace. The task is solved using only a simple location reward (Eq.~\ref{eq:r_location}), without additional reward shaping or engineering. Despite this minimal design, the character reaches targets naturally and robustly, exhibiting seamless whole-body coordination such as stopping, squatting, and side-stepping when needed.


\subsubsection{Velocity Control.}
We consider a velocity control task where the agent is required to track time-varying target speeds and headings. We define the character velocity $\mathbf{v}_{\text{joint}}$ as the linear velocity of the pelvis joint projected onto the horizontal plane. The velocity reward $R_{\text{velocity}}$ is formulated as a weighted sum of a velocity tracking term and an orientation alignment term; 

\begin{equation}
\label{eq:r_velocity}
R_{\text{velocity}} = (1-\lambda) \, \cdot \, \exp\left(-\alpha_3 \, \cdot \, \lVert \mathbf{v}_{\text{target}} - \mathbf{v}_{\text{joint}} \rVert^2\right)
+ \lambda \, \cdot \, \frac{\mathbf{d}_{\text{fwd}} \cdot \mathbf{d}_{\text{target}} + 1}{2}
\end{equation}
where $\mathbf{d}_{\text{fwd}}$	and $\mathbf{d}_{\text{target}}$ represent the current and target heading vectors, respectively. This formulation ensures that the character not only reaches the target speed but also aligns its body orientation with the direction of travel for more natural movement, as demonstrated in Fig.~\ref{fig:teaser}(c).

\subsubsection{Object Interaction.}
To evaluate our framework’s ability to handle contact-rich interactions, we introduce a sofa-sitting task, shown in Fig.~\ref{fig:teaser}(d). The agent is required to lower its center of mass onto a target seating region on the sofa, whose position and yaw are randomized at initialization to encourage generalization. The reward follows the same formulation as $R_{\text{location}}$, penalizing the distance between the character’s pelvis and the target seating region.

\subsubsection{Adaptation on Uneven Terrain.}
We hypothesize that, although the pretrained diffusion prior may fail to maintain balance on uneven terrain, its motion manifold still contains latent primitives, \eg, high-stepping and balance recovery, that can be leveraged for locomotion in challenging environments. To test this, we evaluate our policies on uneven terrains, which are unseen in the diffusion prior's training data. Unlike loss-based test-time guidance, which would require an explicit analytical model of complex terrain geometry, our approach simply augments the policy observation with a terrain height map. The height map encodes terrain elevations within a $4\times4$~meter square centered on the agent’s head at a resolution of 12.5 cm.

To facilitate stable locomotion on irregular surfaces, we employ curriculum learning~\cite{bengio2009curriculum} for both far goal reaching and agile hand reaching (Fig.~\ref{fig:teaser}(a),(c)). Training starts on relatively flat terrain with nearby targets. Once the mean reward exceeds a threshold, we progressively increase both terrain difficulty, \eg, steeper stairs, rougher slopes, discrete blocks, and target distance. Please refer to the supplementary materials for more details.

\section{Experiments and Results}
\subsection{Experiment Setup}

\subsubsection{Implementation Details.}

When pre-training the diffusion prior, we follow the training paradigm of UniPhys \cite{uniphys}, utilizing a 12-layer causal Transformer decoder with a 768-dimensional hidden state and 8 attention heads. The model takes 224-dimensional compact state representations alongside 32-dimensional latent action embeddings as input. When training the noise navigation policy with PPO, we freeze the pretrained diffusion prior and use 5-step DDIM denoising. To balance reactivity and efficiency, the action chunk size $k$ is dynamically adjusted based on task demands: $k=4$ is utilized for uneven terrain and agile reaching tasks to enhance near closed-loop control, while $k=8$ is used for flat-ground scenarios to maximize inference speed. Within each chunk, the pipeline outputs target joint positions $\mathbf{a}_t \in \mathbb{R}^{J \times 3}$, which are converted to torques via a proportional-derivative (PD) controller. These torques drive the transition to the next state $\mathbf{s}_{t+1} = \mathcal{SIM}(\mathbf{s}_t, \mathbf{a}_t)$ within the Isaac Gym simulator \cite{isaacgym}. The simulation environment employs a 30Hz control frequency and an SMPL-like character \cite{smpl} with 24 rotational joints. We adopt an early termination criterion if the pelvis height drops below 0.15 m \cite{deepmimic}. More PPO training details and task-specific reward settings are available in the supplementary material.


\subsubsection{Baselines.}
We evaluate our method against four representative physics-based character control baselines. (1) UniPhys \cite{uniphys}, a unified diffusion-based policy that relies on gradient-based test-time guidance for task-specific control;
(2) PULSE \cite{pulse}, a per-task RL policy utilizing a VAE prior; 
(3) CLoSD \cite{closd}, a hierarchical framework that decouples control into a kinematic diffusion planner and an RL-based low-level tracker; 
and (4) MaskedMimic \cite{maskedmimic}, a unified RL controller that distills a multi-task policy using a masked cVAE. In addition to these general controllers, we evaluate against three representative task-specific RL baselines utilizing adversarial motion priors for the far goal-reaching task:(5) AMP \cite{2021-TOG-AMP}, (6) CML \cite{composite}, and (7) AdaptNet \cite{adaptnet}. We used the officially released models for all baselines, except for PULSE and AMP, which we trained with the same reward as our method.


\subsubsection{Evaluation Metrics.}
For all tasks, we measure task success rates and error metrics across 1000 episodes to evaluate the policy performance. We further assess physical plausibility by quantifying motion smoothness via jerk, the time derivative of acceleration. Fallen cases are excluded from these plausibility measurements. We also benchmark the inference efficiency (frames per second, FPS) against the diffusion-based baseline. To evaluate sample efficiency, we report the total number of Markov Decision Process (MDP) steps in Table~\ref{tab:task_specific_baselines}, where each step corresponds to an action segment spanning 4 frames for terrain, and 8 frames for flat ground. Sample counts for AMP are obtained from our own training runs, while metrics for CML and AdaptNet are sourced directly from their original papers. More qualitative visual results can be found in the supplementary video. 
\begin{table}[t]
  \centering
  \scriptsize
  \setlength{\tabcolsep}{3pt}
  \renewcommand{\arraystretch}{1.15}
  \caption{Comparison with task-specific RL baselines on the far-goal reaching task.}
  \label{tab:task_specific_baselines}
    \begin{tabular}{@{}lcccc@{}}
      \toprule
      Type & Methods & Jerk $(m/s^3)\downarrow$ & Success Rate $\uparrow$ & Samples $\downarrow$ \\ [0.5ex]
      \midrule
      \multirow{3}{*}{Flat}
      & AMP \citesupp{2021-TOG-AMP} & 2387 & 0.967 & 1200M \\
      & CML \citesupp{composite} & 2185 & 0.963 & 400M \\
      & \textbf{NaP(Ours)} & \textbf{1266} & \textbf{0.984} & 327M \\
      \midrule
      \multirow{2}{*}{Terrain} 
      & AdaptNet \citesupp{adaptnet} & 2444 & \textbf{0.868} & 408M \\
      & \textbf{NaP(Ours)} & \textbf{1788} & 0.860 & 900M \\
      \bottomrule
    \end{tabular}%
\end{table}

\subsection{Far Goal Reaching}
This task evaluates long-horizon stability through far-goal reaching. During training, pelvis goals are set within a 0.9–0.92 m height and a 6 m horizontal radius. During evaluation, goals are fixed at a 5 m distance and uniformly sampled within a $\pm 45^{\circ}$ sector relative to the character’s forward heading. Success requires the pelvis to stay within 0.3 m of the goal for 0.5 s. 

As presented in Table~\ref{tab:inference_speed} and Table~\ref{tab:naturalness_flat}, compared with the diffusion-based UniPhys, which depends on test-time guidance to reach the goal, NaP achieves both substantially faster inference and higher success: 22.5 FPS vs. 2.9 FPS and 98.4\% vs. 81.9\%, while still preserving motion smoothness and naturalness. Against RL-based baselines (CLoSD and PULSE), NaP attains a similarly high success rate but produces significantly smoother and more natural motions. In particular, PULSE exhibits noticeable high-frequency jitter and less fluid transitions, whereas CLoSD often yields abrupt turning behaviors and severe jittering, especially in the hand and foot joints. MaskedMimic, in contrast, fails to reach distant goals robustly. Similarly, compared to task-specific adversarial baselines, NaP attains higher motion naturalness than AMP as shown in Table~\ref{tab:task_specific_baselines} and delivers superior motion quality over CML and AdaptNet while maintaining a comparable success rate. We also compare against a transformer-based task-specific RL baseline on a path-following task. Please refer to the supplementary material for details.

These results demonstrate that navigating latent noise selection in a pretrained diffusion-based action prior using task rewards and environmental feedback enables robust, goal-directed behavior while preserving motion realism. Importantly, NaP removes the need for expensive test-time optimization, as the policy directly learns to generate physically feasible and task-successful actions through efficient noise navigation. Additional qualitative comparisons are provided in the supplementary video.

\begin{table}[tb]
  \caption{Inference efficiency comparison with diffusion-based baseline.}
  \label{tab:inference_speed}
  \centering
  \scriptsize
  \setlength{\tabcolsep}{4pt}
  \renewcommand{\arraystretch}{1.05}
  \begin{tabular}{@{}lcc@{}}
    \toprule
    Inference Speed (FPS) $\uparrow$ & Far Goal Reaching & Velocity Control \\
    \midrule
    UniPhys & 2.9 & 5.2 \\
    NaP-Control (Ours) & 22.5 & 24.4 \\
    \bottomrule
  \end{tabular}
\end{table}

\begin{table}[tb]
  \caption{Flat-ground motion naturalness and success rate comparison.}
  \label{tab:naturalness_flat}
  \centering
  \setlength{\tabcolsep}{3pt}
  \renewcommand{\arraystretch}{1.15}

  \resizebox{\linewidth}{!}{%
    \begin{tabular}{@{}lccccccccc@{}}
      \toprule
      \multirow{2}{*}{Methods} &
      \multicolumn{2}{c}{Far Goal Reaching} &
      \multicolumn{2}{c}{Agile Hand Reaching} &
      \multicolumn{3}{c}{Velocity Control} &
      \multicolumn{2}{c}{Object Interaction} \\
      \cmidrule(lr){2-3} \cmidrule(lr){4-5} \cmidrule(lr){6-8} \cmidrule(lr){9-10}
      & \shortstack{Jerk $\downarrow$\\$(m/s^3)$} & \shortstack{Success Rate $\uparrow$\\~} &
        \shortstack{Jerk $\downarrow$\\$(m/s^3)$} & \shortstack{Success Rate $\uparrow$\\~} &
        \shortstack{Jerk $\downarrow$\\$(m/s^3)$} & \shortstack{Velocity Error $\downarrow$\\$(m/s)$} & \shortstack{Success Rate $\uparrow$\\~} &
        \shortstack{Jerk $\downarrow$\\$(m/s^3)$} & \shortstack{Success Rate $\uparrow$\\~} \\
      \midrule
      PULSE \cite{pulse} & 2594 & \textbf{0.998} & 3631 & \textbf{0.998} & 2273 & \textbf{0.0455} & \textbf{0.996} & 1373 & \textbf{0.999} \\
      MaskedMimic \cite{maskedmimic} & 3564 & 0.304 & 2750 & 0.725 & 2420 & \underline{0.0678} & 0.355 & 1310 & 0.298\\
      CLoSD \cite{closd} & 1876 & \textbf{0.998} & -- & -- & -- & -- & -- & 1505 & 0.862 \\
      UniPhys \cite{uniphys} & \textbf{528} & 0.819 & -- & -- & \textbf{401} & 0.7971 & 0.921 & -- & -- \\
      \textbf{NaP-Control(Ours)}& \underline{1266} & \underline{0.984} & \textbf{1146} & \underline{0.991} & \underline{752} & 0.1753 & \underline{0.990} & \textbf{491} & \underline{0.979} \\
      \bottomrule
    \end{tabular}%
  }
\end{table}

\subsection{Agile Hand Goal Reaching}
This task evaluates motion agility and reactivity through right-hand reaching with frequently changing targets sampled from all directions. During evaluation, targets are placed within a radius of 1 m, with heights ranging from 0.4 to 2.0 m. The task requires rapid postural adaptation while maintaining temporal smoothness and stability. A trial is considered successful if the hand remains within 0.3 m of the target for at least 0.2 s.

UniPhys and CLoSD fail to complete this task due to insufficient fine-grained control, while MaskedMimic achieves a relatively low success rate. Consistent with the far-goal reaching results, NaP achieves a success rate comparable to PULSE, but with substantially reduced motion jitter (jerk 1146 $m/s^3$ vs. 3631 $m/s^3$) and noticeably smoother transitions. Compared to far goal reaching, the advantage in motion naturalness becomes even more pronounced in this setting, as the task demands rapid directional and height changes that strongly challenge the temporal coherence of conventional RL-based policies.

\subsection{Velocity Control}
For the velocity control task, the direction and speed of the target velocity are randomly sampled within 3$m/s$ for each episode. We allow a 90-frame (3s) transition period to reach the target velocity before calculating the mean tracking error, while motion smoothness is quantified over the entire 200-frame (7s) non-falling episodes to capture the quality of the initial acceleration and facing-direction transition phase. The success rate is defined as the ratio of episodes in which the character completes the motion without falling to the total number of episodes. CLoSD does not provide velocity control, and MaskedMimic fails the task with a low success rate. PULSE achieves a low velocity error, but produces unnatural running poses with a high jerk. In contrast, UniPhys produces much smoother motion with a jerk of 401 $m/s^3$, but at the cost of a significant velocity error of 0.7971 $m/s$. NaP effectively bridges this gap, generating smooth motion with a jerk of 752 $m/s^3$, which is comparable to the performance of UniPhys, while achieving a velocity error of 0.1753 $m/s$.

\subsection{Object Interaction}
\begin{figure}[t]
  \centering
  \includegraphics[width=\linewidth]{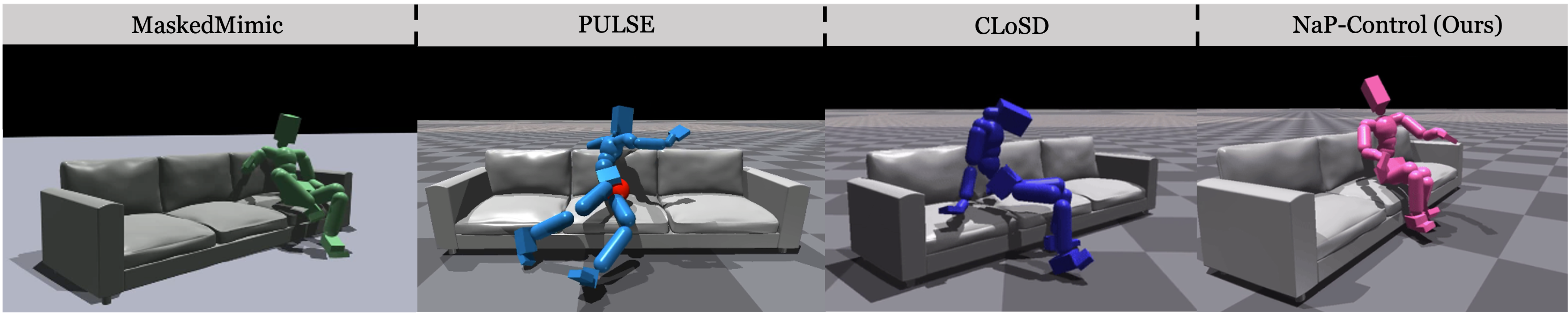}
  \caption{Qualitative Comparison of Object Interaction Task.}
  \label{fig:sitting}
\end{figure}

We evaluate the agent's ability to reach a target sitting position on a sofa. Success is achieved when the pelvis is within 0.15m horizontally and 0.1m vertically of the target height. Although our diffusion prior was not trained on human-object interaction data, NaP can identify the most suitable motion for the sitting task while maintaining natural sitting motions. In contrast, CLoSD and MaskedMimic have a low task success rate; UniPhys cannot perform the task; PULSE, without a specific reward for human-like motion, produces highly unnatural behaviors to satisfy the goal, such as severe upper-body twisting (see Fig.~\ref{fig:sitting}). This comparison underscores that NaP inherently preserves the manifold of natural human movement, even when applied to object interaction scenarios for which no explicit data exist in the training of the diffusion prior.
\subsection{Performance on Uneven Terrain}
By incorporating terrain height maps, we extend our framework to uneven terrains, demonstrating strong generalization to previously unseen environments. We note that diffusion-based baselines such as UniPhys and CLoSD are trained and evaluated on flat-ground settings in their original formulations, and therefore cannot be directly deployed on uneven terrains. Consequently, we restrict our benchmarking to PULSE and MaskedMimic. Consistent with our flat-ground results, NaP achieves goal-reaching success rates and velocity tracking errors comparable to PULSE, and significantly outperforms MaskedMimic, while producing substantially smoother and more human-like motion. These results highlight that navigating latent noise selection within a pretrained diffusion-based action prior enables robust adaptation to new environments, allowing our approach to extend naturally to more challenging and previously unseen scenarios. Quantitative comparisons are summarized in Table \ref{tab:naturalness_uneven}.

\begin{table}[tb]
  \caption{Uneven-terrain motion naturalness and success rate comparison. While MaskedMimic minimizes velocity error, it suffers from a lower success rate.}
  \label{tab:naturalness_uneven}
  \centering
  \setlength{\tabcolsep}{3pt}
  \renewcommand{\arraystretch}{1.15}

  \resizebox{\linewidth}{!}{%
    \begin{tabular}{@{}lcccccccccc@{}}
      \toprule
      \multirow{2}{*}{Methods} &
      \multicolumn{2}{c}{Far Goal Reaching} &
      \multicolumn{2}{c}{Agile Hand Reaching} &
      \multicolumn{3}{c}{Velocity Control} \\
      \cmidrule(lr){2-3} \cmidrule(lr){4-5} \cmidrule(lr){6-8}
      & \shortstack{Jerk $(m/s^3)$$\downarrow$} & \shortstack{Success Rate $\uparrow$} &
        \shortstack{Jerk $(m/s^3)$$\downarrow$} & \shortstack{Success Rate $\uparrow$} &
        \shortstack{Jerk $(m/s^3)$$\downarrow$} & \shortstack{Velocity Error $(m/s)$$\downarrow$} & \shortstack{Success Rate $\uparrow$} \\
      \midrule
      PULSE \cite{pulse} & 2830 &  \textbf{0.966} & 3976 & \textbf{0.980} & 2397 & \underline{0.0569} & \textbf{0.995} \\
      MaskedMimic \cite{maskedmimic} & 3566 & 0.319 & 2723 & 0.676 & 2262 & \textbf{0.0556} & 0.292\\
      \textbf{NaP-Control(Ours)} & \textbf{1788} & \underline{0.860} & \textbf{1520}& \underline{0.909}& \textbf{1358} & 0.1654 & \underline{0.950}\\
      \bottomrule
    \end{tabular}%
  }

\end{table}

%

\subsection{Ablation Studies}


\subsubsection{Motion State Representation.} We adopt a compact yet informative state representation that includes global root trajectories, joint rotations, and angular velocities, while excluding local joint positions and linear velocities. This design reduces the effective search space without sacrificing key motion cues, balancing RL exploration efficiency and control stability, as evidenced by our flat-ground far goal-reaching results (Fig.~\ref{fig:ablation}(a)). Using this compact representation yields more efficient exploration than the redundant full state used in UniPhys and substantially outperforms a minimal state that retains only root-related features.

\subsubsection{Action Chunking Size.} While action chunks of 4–8 steps generally balance temporal consistency and reactive control (Fig.~\ref{fig:ablation}(b)), task difficulty dictates the ideal granularity. As shown in Fig.~\ref{fig:ablation}(c), smaller chunk sizes favor stability on uneven terrain and achieving precision in agile hand-reaching tasks.

\subsubsection{Noise Navigation Space.} 
\begin{figure*}[t]
  \centering
  \includegraphics[width=0.25\textwidth]{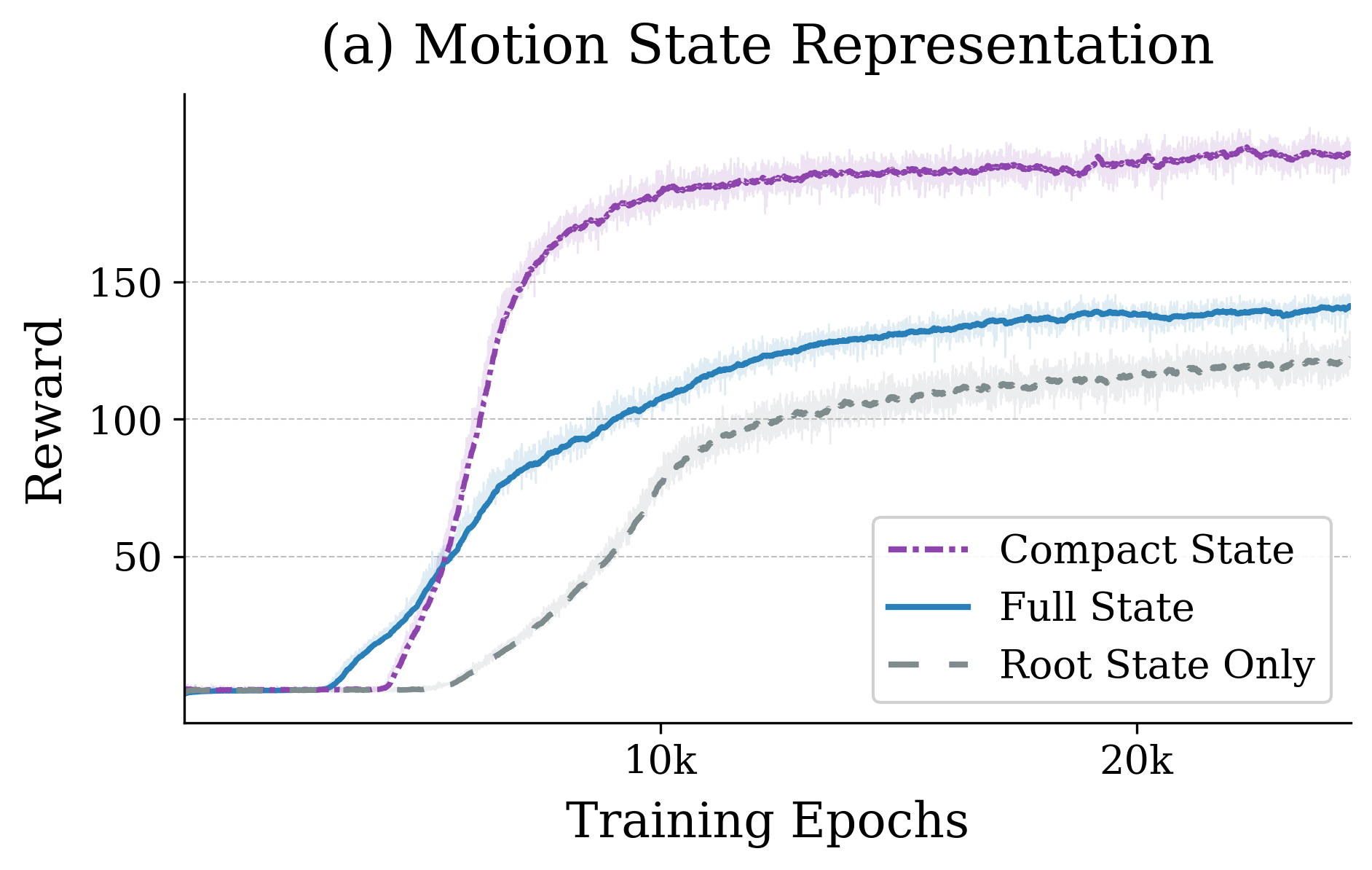}\hfill
  \includegraphics[width=0.25\textwidth]{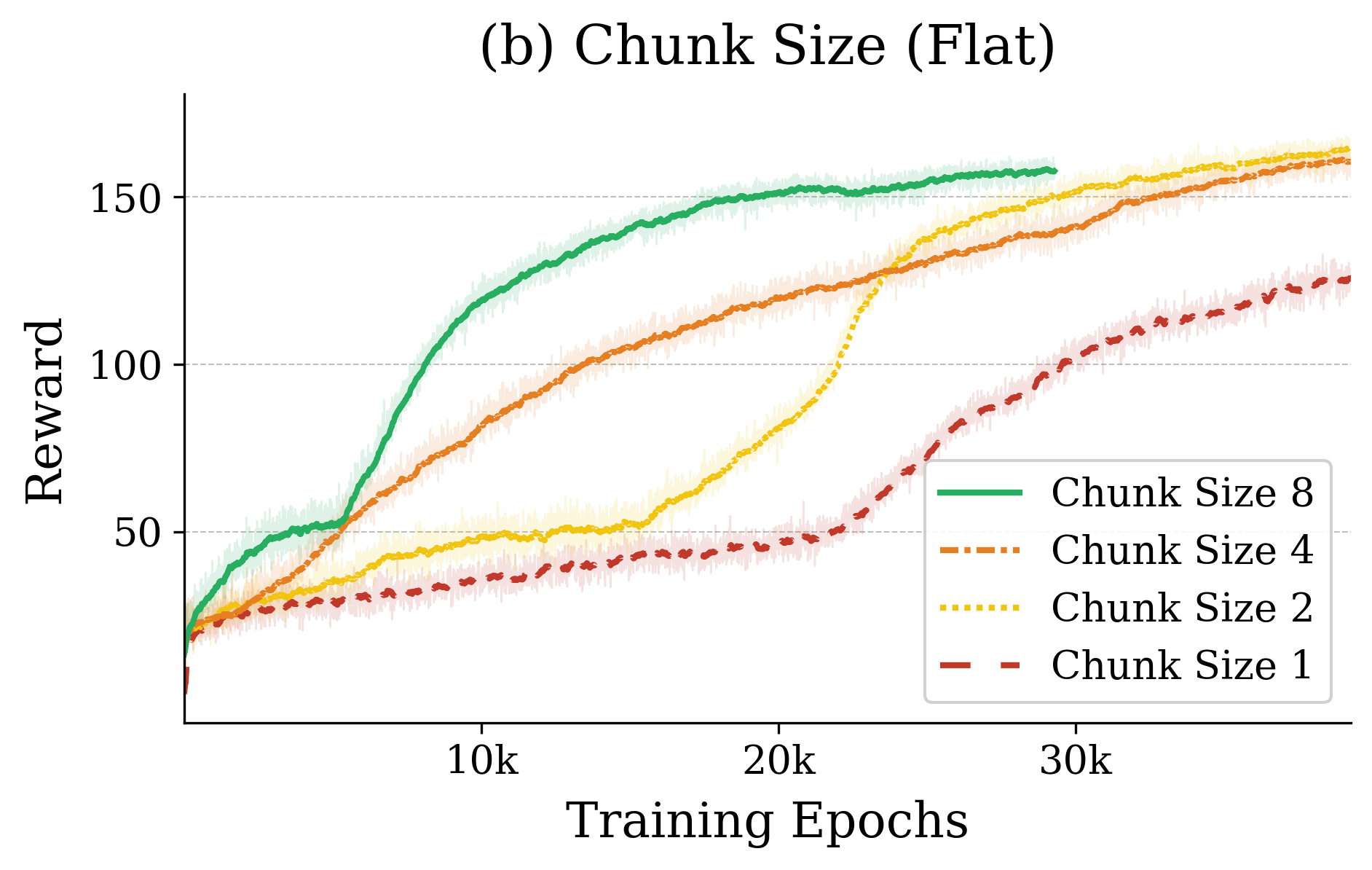}\hfill
  \includegraphics[width=0.25\textwidth]{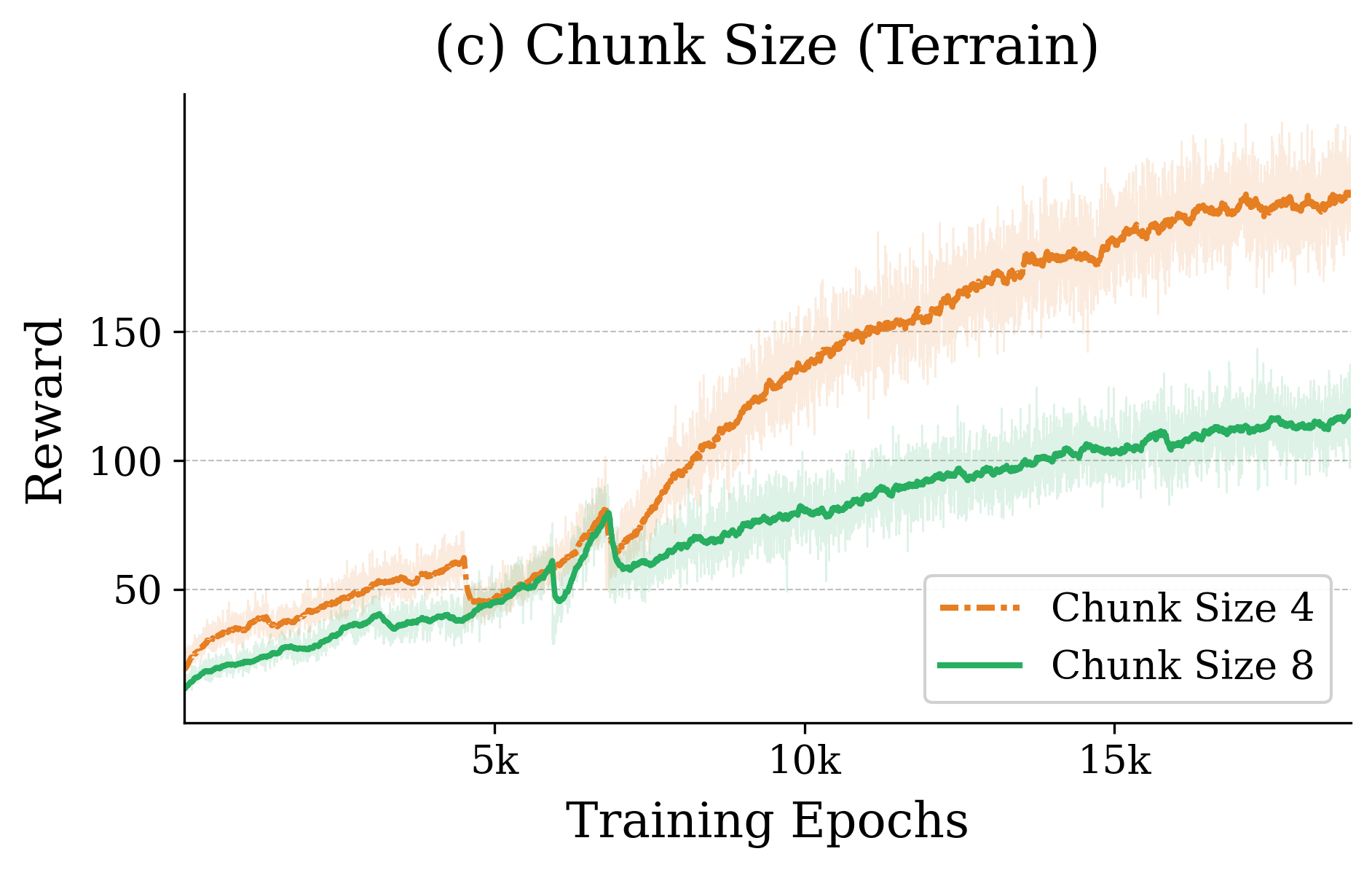}\hfill
  \includegraphics[width=0.25\textwidth]
  {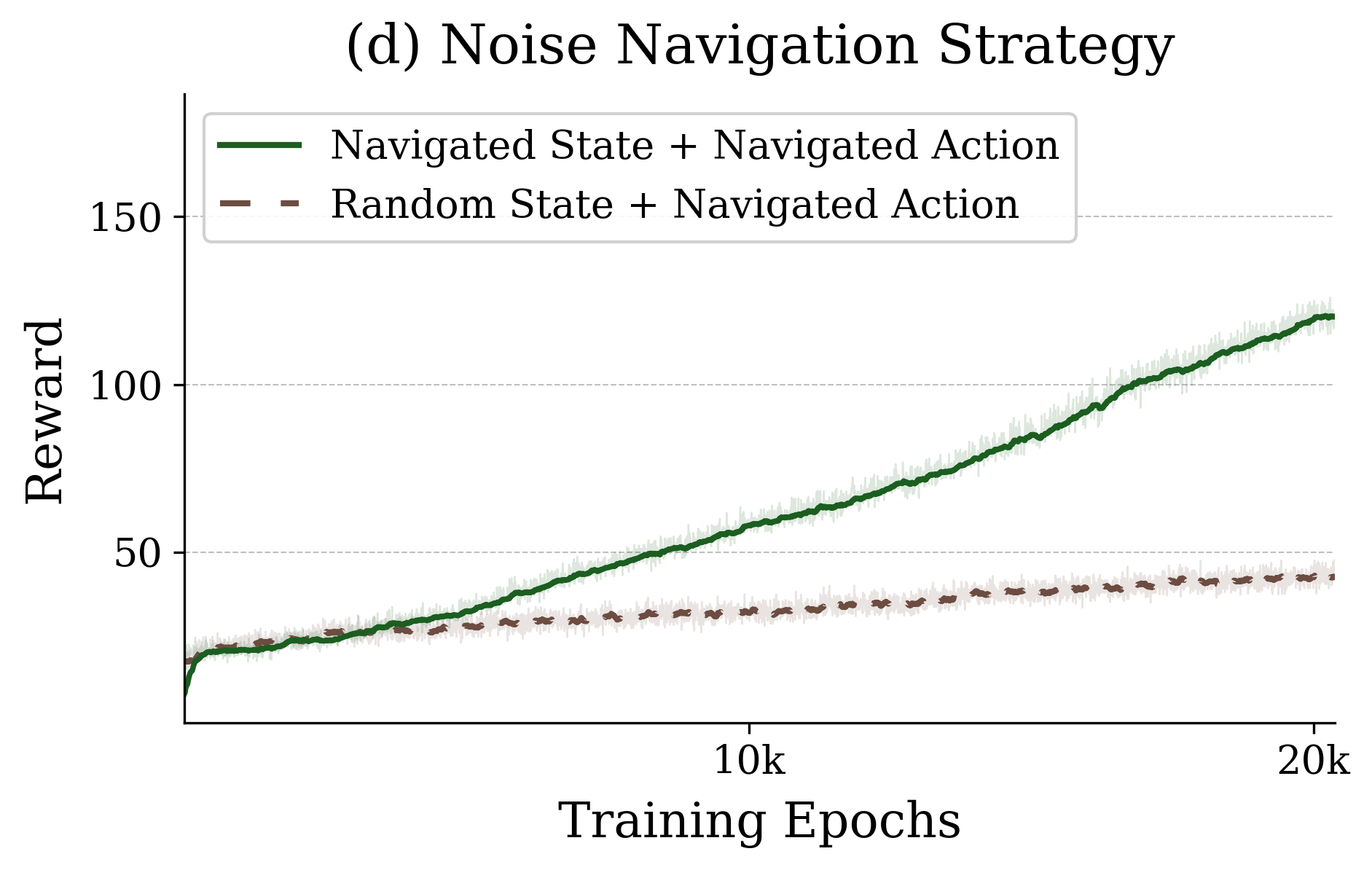}
  \caption{Ablation studies. (a) Effect of state representation on flat-ground far goal reaching. (b--c) Effect of action chunk size $k$ for agile hand reaching on flat ground (b) and uneven terrain (c). (d) Comparison of joint state-action noise optimizing versus action-only noise optimizing for agile hand reaching.}
  \label{fig:ablation}
\end{figure*}

While task execution only requires action output, we find that navigating both state and action noise spaces significantly outperforms navigating action noise space alone. As evaluated on flat-ground agile hand reaching (Fig.~\ref{fig:ablation}(d)), this joint strategy provides the RL policy with finer control over the diffusion prior's generative manifold, resulting in more robust and physically plausible task completion.

\begin{figure}[t]
  \centering
  \includegraphics[width=0.5\linewidth]{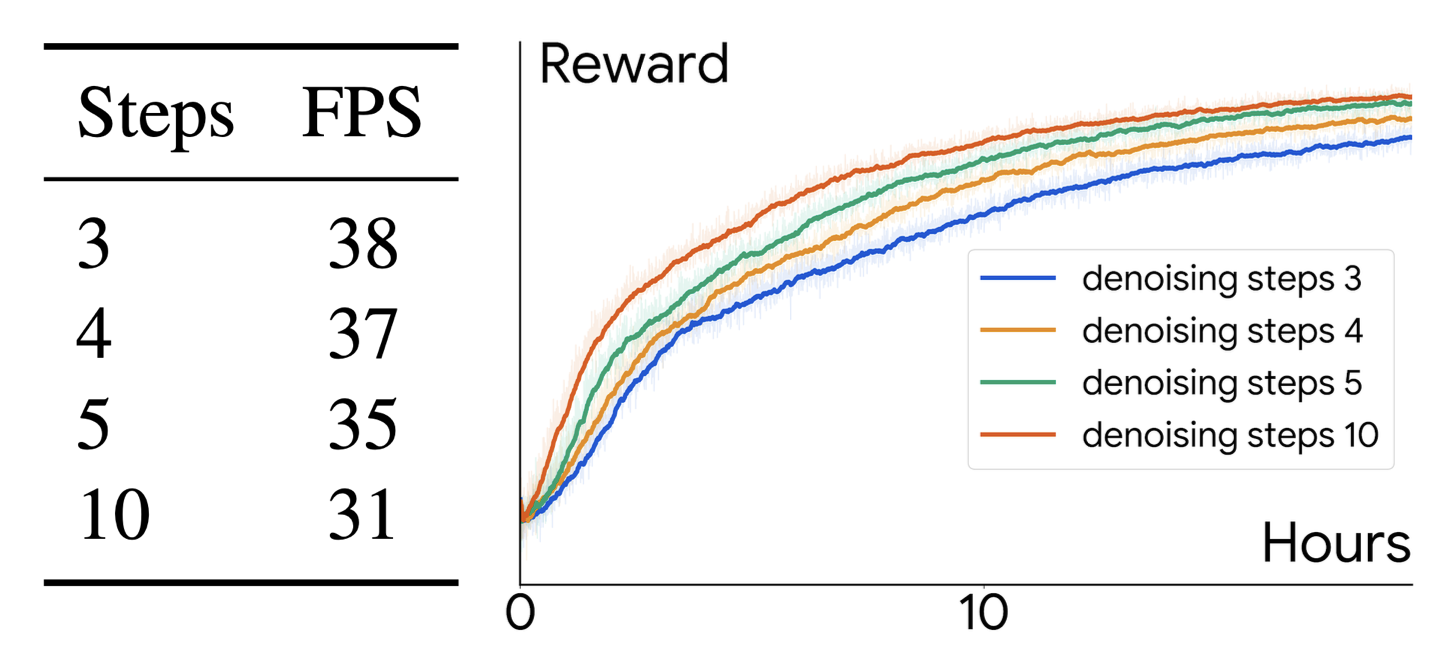}
  \caption{Training convergence of the velocity control task on flat ground with different diffusion denoising steps.}
  \label{fig:denoise_steps_training}
\end{figure}
\subsubsection{ODE-Solver Step and Inference Speed.}
\label{sec:inference-speed}
We evaluate the inference speed across varying diffusion denoising steps for the velocity control task on flat ground. 5 denoising steps are a practical trade-off between training and inference efficiency as shown in Fig.~\ref{fig:denoise_steps_training}. Inference can be accelerated by reducing the number of denoising steps, at the cost of increased training effort.
\section{Conclusion}



We presented NaP-Control, a framework for fast and robust whole-body character control that navigates the latent noise of a task-agnostic diffusion policy toward task-specific behaviors. By training a compact RL agent to directly predict the diffusion noise that leads to high-reward actions, NaP preserves the motion quality of the prior while eliminating both the computational overhead and fragility of gradient-based control at inference. A key property of NaP is that it learns by interacting directly with the environment, enabling it to generate motion under complex dynamics and handle non-differentiable task objectives, which enhances robustness in difficult conditions. Consequently, NaP achieves higher success rates with low-latency inference, making diffusion-based control viable for real-time use. We believe NaP provides a foundation for generating natural, expressive motion in diverse tasks that require long-horizon stability and precise responsiveness, including whole-body loco-manipulation, contact-intensive interactions, and dynamic multi-agent settings.

%
\bibliographystyle{splncs04}
\bibliography{main}
\clearpage
\setcounter{page}{1}
\pagenumbering{arabic}

\setcounter{section}{0} 
\section{Supplementary}
We provide comprehensive qualitative results and side-by-side baseline comparisons in the accompanying supplementary video (supp\_video.mp4). We strongly encourage readers to view this video for a detailed demonstration of our framework.

In addition, this supplementary document provides technical specifications and implementation details to complement the main manuscript. We include comprehensive definitions for all task-specific symbols, hyperparameter configurations for our latent noise navigation policy, and a detailed breakdown of the motion state representations used in our ablation study. Finally, we discuss the limitations of our current framework and propose potential avenues for future work.

\subsection{Task Specifications and Reward Design}
This section provides detailed notations, symbol definitions, and hyperparameter values to complement the reward formulations described in the main manuscript. Table~\ref{tab:task_inputs} specifies the policy inputs for each task, while Table~\ref{tab:task_symbols} defines the symbols used in the reward functions (Eq.3–7). For the far goal reaching task, stability penalties for linear and angular velocities are omitted when the pelvis's motion falls below the thresholds $v_{threshold}$ and $w_{threshold}$, respectively. This thresholding mechanism ensures that the stability terms do not over-penalize minor fluctuations at low speeds. Furthermore, we incorporate sparse rewards upon task completion for both far goal reaching and agile hand reaching. These rewards incentivize the agent to satisfy success criteria directly, preventing suboptimal behaviors such as circling the target to accumulate dense rewards.

\begin{table}[h]
  \caption{Policy inputs for each task.}
  \label{tab:task_inputs}
  \centering
  \small
  \setlength{\tabcolsep}{4pt}
  \renewcommand{\arraystretch}{1.15}
  \begin{tabular}{@{}ll@{}}
    \toprule
    \textbf{Task} & \textbf{Task-specific input} \\
    \midrule
    Far goal reaching & Goal coordinates; terrain height maps (optional) \\
    Agile hand reaching & Goal coordinates; terrain height maps (optional) \\
    Velocity control & Target speed and direction; terrain height maps (optional) \\
    Object interaction & Goal coordinates; sofa states \\
    \midrule
    Shared input & Character's proprioception \\
    \bottomrule
  \end{tabular}
\end{table}

\begin{table}[h]
  \caption{Task-specific symbols used in task setup and reward design.}
  \label{tab:task_symbols}
  \centering
  \small
  \setlength{\tabcolsep}{4pt}
  \renewcommand{\arraystretch}{1.15}
  \resizebox{\linewidth}{!}{%
  \begin{tabular}{@{}llr@{}}
    \toprule
    \textbf{Symbol} & \textbf{Description} & \textbf{Value / range} \\
    \midrule
    \multicolumn{3}{@{}l}{\textbf{Far goal reaching}} \\
    $p_{\text{goal}}$ & Target goal position in world coordinates. &  \\
    $p_{\text{joint}}$ & Pelvis position in world coordinates. &  \\
    $\mathbf{d}_{\text{fwd}}$ & Character forward heading unit vector. &  \\
    $\mathbf{d}_{\text{goal}}$ & Unit vector pointing from $p_{\text{joint}}$ to $p_{\text{goal}}$. & \\
    $\alpha_1,\alpha_2$ & Scaling factors for $R_{\text{location}}$ and $R_{\text{orientation}}$. & 4.0 / 2.0 (uneven), 2.0 \\
    $\lambda_1,\lambda_2$ & Weights for orientation and stability rewards in Eq.3. & 0.05, 0.35 / 0.35, 0.35 (uneven) \\
    $v$ & Linear speed magnitude used in stability penalty in Eq.6. &  \\
    $\omega$ & Angular speed magnitude used in stability penalty in Eq.6. &  \\
    $c_1,c_2$ & Coefficients combining linear and angular speeds in Eq.6. & 0.05, 0.15 \\
    $\beta_1,\beta_2$ & Stability penalty scales in Eq.6. & 2.0, 1.0 \\
    $v_\text{threshold}$ & Velocity threshold, $c_1$ is 0 if $v \leq v_\text{threshold}$. & 0.6 \\
    $w_\text{threshold}$ & Angular velocity threshold, $c_2$ is 0 if $w \leq w_\text{threshold}$. & 0.5 \\
    $R_\text{sparse}$ & Sparse reward when suceess. & 100 \\
    \addlinespace[2pt]
    \multicolumn{3}{@{}l}{\textbf{Agile hand reaching}} \\
    $p_{\text{goal}}$ & Target hand position in world coordinates. &  \\
    $p_{\text{joint}}$ & Right-hand end-effector position in world coordinates. &  \\
    $\alpha_1$ & Scaling factors for $R_{\text{location}}$. & 2.0 \\
    $R_\text{sparse}$ & Sparse reward when success. & 100 \\
    \addlinespace[2pt]
    \multicolumn{3}{@{}l}{\textbf{Velocity control}} \\
    $\mathbf{v}_{\text{target}}$ & Target horizontal velocity vector. &  \\
    $\mathbf{v}_{\text{joint}}$ & Pelvis linear velocity projected onto horizontal plane. &  \\
    $\mathbf{d}_{\text{target}}$ & Target heading unit vector (direction of $\mathbf{v}_{\text{target}}$). &  \\
    $d_{\text{fwd}}$ & Character forward heading unit vector. &  \\  
    $\alpha_3$ & Scaling factor for velocity tracking in Eq.7. & 0.25 \\
    $e$ & Orientation error. & $1 - \frac{\mathbf{d}_{\text{fwd}} \cdot \mathbf{d}_{\text{target}} + 1}{2}$ \\
    $\lambda$ & Weight balancing tracking vs. heading alignment in Eq.7. &  $\lambda = \frac{e}{e + 0.05}$ \\
    \addlinespace[2pt]
    \multicolumn{3}{@{}l}{\textbf{Object interaction}} \\
    $p_{\text{goal}}$ & Target seating region position on the sofa. &  \\
    $p_{\text{joint}}$ & Pelvis position. &  \\
    $\alpha_1$ & Scaling factors for $R_{\text{location}}$. & 15.0 \\
    &     The sofa is positioned at a distance of 0.95 to 1.05 m behind the character,& \\
    &     with a 30-degree yaw rotation applied. & \\
    \addlinespace[2pt]
    \multicolumn{3}{@{}l}{\textbf{Uneven terrain adaptation (optional input)}} \\
    $H$ & Terrain height map centered on the agent. & $4\times4$ m spatial window, 12.5 cm resolution \\
    $k$ & Action chunk length used for open-loop execution. & $k{=}8$ (flat), $k{=}4$ (uneven, agile hand reaching) \\
    \multicolumn{3}{@{}l}{\textbf{Diffusion Prior Sampling}} \\
    $\mathcal{H}$ & History buffer of character states and latent actions &  \\
     & (Initialized using the mean and standard deviation of the training dataset.) &  \\
    $h$ & History buffer size. & 4 \\
    \bottomrule
  \end{tabular}}
\end{table}

\subsection{Implementation Details}
We provide the full set of hyperparameters for our PPO-based latent noise navigation policy in Table \ref{tab:implementation_details}. This includes the network architectures for both the Base MLP and the Terrain MLP, as well as the specific curriculum levels used to train the agent on uneven surfaces. For flat-ground tasks, the Base MLP processes both the proprioceptive and task-specific environmental states. In contrast, for uneven-terrain tasks, the Base MLP is restricted to proprioceptive inputs, while environmental states, including terrain height maps, are encoded separately via the Task MLP. In our implementation, the total loss function (Supp Eq.~\ref{eq:loss_total}) incorporates a standard clipped surrogate objective $L^{CLIP}$ \citesupp{ppo}, a value function squared-error loss $L^{VF}$ to refine the critic's state-value predictions, and the boundary loss $L^{B}$ to penalize the policy's mean output $\mu$ if it exceeds a specified threshold ($\pm 1.0$):

\begin{equation}
\label{eq:loss_total}
    L(\theta) = L^{CLIP}(\theta) + b_1 L^{VF}(\theta) + b_2 L^{B}(\mu)
\end{equation}

\begin{table}[h]
    \centering
    \caption{Implementation details of the latent noise navigation policy. $^{*}$ For velocity control task on uneven terrains.}
    \label{tab:implementation_details}
    \small
    \begin{tabular}{@{}ll@{}}
    \toprule
    \textbf{Category} & \textbf{Value / Configuration} \\ \midrule
    \textbf{Training Infrastructure} & \\
    Hardware & RTX 4090 GPU, 4 CPUs, 32GB RAM \\
    $^{*}$Hardware & NVIDIA H200, 141GB RAM\\
    Parallel Environments & 256 / 1536$^{*}$\\ 
    \textbf{Inference Infrastructure} & RTX TITAN GPU \\ \midrule
    \textbf{PPO Hyperparameters} & \\
    Learning Rate & $2 \times 10^{-5}$ (Constant) \\
    Discount Factor ($\gamma$) & 0.99 \\
    GAE Parameter ($\tau$) & 0.95 \\
    Clip Range ($\epsilon$) & 0.2 \\
    Horizon Length & 32 \\
    Minibatch Size & 4096 / 16384$^{*}$\\
    Mini-epochs & 8 / 6$^{*}$\\
    Critic Loss and Boundary Loss Coefficients & $b_1$: 5, $b_2$: 10 \\ \midrule
    \textbf{Network Architecture} & \\
    Base MLP & 3 layers: [2048, 1024, 512] \\
    Task MLP(only used for the terrain task) & 2 layers: [512, 256] \\
    Activation Function & SiLU \\
    Normalization & Pre-LayerNorm (Only for middle layers) \\ \midrule
    \textbf{Uneven Terrains \& Curriculum} & \\
    Terrain Types & Smooth/Rough Slope, Stairs Up/Down, Discrete \\
    Terrain Proportions & [0.25, 0.15, 0.25, 0.25, 0.1] \\
    Terrain Levels & 5 \\
    Curriculum Thresholds (for goal reaching) & [50, 100] (unlock level 1; unlock all) \\ \midrule
    \textbf{Constraints} & \\
    Action Bounds & None (Enforced via boundary loss $L^{B}$) \\
    Discriminator Loss & None \\ \bottomrule
    \end{tabular}
\end{table}

In the ablation study of the main manuscript, we evaluated the performance impact of different state representations for the diffusion prior. Table~\ref{tab:state_representations} provides the exact decomposition of the features for each variant compared. Our evaluation in the main manuscript identifies the Compact State as the most effective configuration, offering a streamlined representation that maintains motion stability while improving exploration efficiency and the overall performance of policy training. 

\begin{table}[h] 
    \centering 
    \caption{Comparison of state representation variants for the diffusion prior.} \label{tab:state_representations} 
    \begin{tabular}{@{}lcccc@{}} 
    \toprule 
    \textbf{Feature} & \textbf{Full State} & \textbf{Compact State} & \textbf{Root State Only} \\
    \midrule 
    Global Root Trajectories & \checkmark & \checkmark & \checkmark \\
    6D Joint Rotations & \checkmark & \checkmark & -- \\
    Joint Angular Velocities & \checkmark & \checkmark & -- \\
    Local Joint Positions & \checkmark & -- & -- \\
    Local Joint Velocities & \checkmark & -- & -- \\
    \midrule 
    \textbf{Total Dimensions} & \textbf{398} & \textbf{256} & \textbf{47} \\
    \bottomrule 
    \end{tabular} 
\end{table}

\subsection{Comparison with Transformer-based Task-Specific Baselines.}
\label{sec:task-specific-baselines}
%
We also compare against TokenHSI~\citesupp{pan2025tokenhsi}, an adaptation method built over a transformer-based RL policy, to evaluate our framework against an alternative task-specific RL policy. Compared to TokenHSI, which requires the task to be included in the multi-task base policy training, NaP achieves comparable performance on the path-following task on flat ground, as shown in Table~\ref{tab:tokenhsi_comparison} while entirely eliminating the need for base model retraining.

\begin{table}[t]
  \centering
  \small
  \caption{Comparison with TokenHSI on the path-following task. We report motion jerk, success rate, trajectory precision, inference speed, and training samples.}
  \label{tab:tokenhsi_comparison}
    \begin{tabular}{@{}lcccc@{}}
      \toprule
      Methods & Jerk $(m/s^3)\downarrow$ & Success Rate $\uparrow$ & Precision $(m)\downarrow$ & FPS $\uparrow$\\
      \midrule
      \textbf{NaP(Ours)} & 570 & 0.972 & 0.186 & 34.3\\
      TokenHSI \citesupp{pan2025tokenhsi} & 575 & 0.995 & 0.089 & 53.9\\
      \bottomrule
    \end{tabular}%
\end{table}



\subsection{Limitations and Future Work}
While NaP-Control currently uses task-specific reward formulations, the integration of a motion-aware diffusion prior significantly reduces the reliance on extensive manual reward shaping. By leveraging the prior’s learned manifold, we demonstrate that a minimal set of 1 to 3 reward terms is sufficient to achieve high-fidelity motion across various downstream tasks. Although developing a fully task-agnostic policy is beyond the scope of this study, recent advancements in unsupervised reinforcement learning, particularly in the use of intrinsic rewards for downstream adaptation \citesupp{laskin2021urlb, touati2021learning}, suggest promising paths for further generalization. We believe that incorporating such unsupervised objectives could further generalize our framework, enabling multi-task control for complex whole-body characters.

While NaP-Control effectively navigates the physics-based character, its performance is inherently tied to the diversity and quality of the pretrained diffusion prior. Because the policy navigates a fixed latent manifold, it may struggle with motor skills that fall outside the distribution of the prior’s training data. To mitigate this dependency, future iterations could incorporate larger and more diverse motion datasets, including recovery primitives such as get-up motions. Furthermore, we envision a residual reinforcement learning framework \citesupp{sentinel} where the latent noise navigation policy provides the primary latent guidance, while a secondary component applies small corrective joint torques on top of the generated actions. This hybrid approach would allow the character to satisfy precise physical constraints even when the prior is slightly misaligned with the task objectives.

The success of NaP-Control in agile reaching and contact-rich sitting demonstrates its potential for dynamic, physics-based loco-manipulation. A promising future direction is to extend this coordination to interactions involving large-scale external force exchange, such as carrying heavy payloads or pushing carts, where the agent must actively compensate for shifting centers of mass. Furthermore, by incorporating high-resolution latent control for finger joints, NaP-Control could potentially synchronize global postural balance with fine-grained contact adjustments, enabling dexterous manipulation while maintaining robust locomotion.

Beyond individual tasks, the ability of our latent navigation policy to incorporate complex environmental states, such as terrain height maps and object representations, enables a promising extension to multi-agent interaction. By treating other characters as dynamic components of the environment state, we can leverage the diffusion prior to generate natural social behaviors. This would allow for the exploration of collaborative tasks, such as multi-person object transport, or competitive scenarios like sports, where the prior acts as a regularizer to ensure that character interactions remain physically plausible and visually natural.

\begingroup
\renewcommand{\refname}{Supplementary References}
\providecommand{\bibname}{Supplementary References}
\renewcommand{\bibname}{Supplementary References}

\endgroup



\end{document}